\documentclass[12pt]{article}
\usepackage{amsmath,amssymb}
\newcommand{\eqdef}{\stackrel{\text{def}}{=}}
\newcommand{\n}{\nonumber\\}

\newcommand{\ignore}[1]{}
\newcommand{\Romannumeral}[1]{\uppercase\expandafter{\romannumeral#1}}
\newcommand{\I}{\text{\Romannumeral{1}}}
\newcommand{\II}{\text{\Romannumeral{2}}}

\newtheorem{theo}{\bf Theorem}[section]
\newtheorem{conj}[theo]{\bf Conjecture}

\allowdisplaybreaks[4]

\begin{document}

\baselineskip=20pt


\newcommand{\Title}[1]{{\baselineskip=26pt
  \begin{center} \Large \bf #1 \\ \ \\ \end{center}}}
\newcommand{\Author}{\begin{center}
  \large \bf  Ryu Sasaki  \end{center}}
\newcommand{\Address}{\begin{center}
    Faculty of Science, Shinshu University,
     Matsumoto 390-8621, Japan\\
      e-mail: ryu@yukawa.kyoto-u.ac.jp
   \end{center}}
\newcommand{\Accepted}[1]{\begin{center}
  {\large \sf #1}\\ \vspace{1mm}{\small \sf Accepted for Publication}
  \end{center}}

\thispagestyle{empty}

\Title{Confining non-analytic exponential potential                     %
 $V(x)= g^2\exp\,(2|x|)$ and its exact Bessel-function solvability}

\Author

\Address
\vspace{1cm}

\begin{abstract}
In a previous paper we have shown that Schr\"{o}dinger equation with the non-analytic attractive exponential
potential $V(x)= -g^2\exp (-|x|)$ is exactly solvable.
It has finitely many discrete eigenstates described by the Bessel function of the first kind $J_{\nu}(z)$
and the eigenvalues are specified by the positive zeros of $J_{\nu}(g)$ and $J'_{\nu}(g)$ as a function of the order $\nu$ with fixed $g>0$.
Now we show the corresponding results for the {\em confining\/} non-analytic  exponential
potential $V(x)= g^2\exp (2|x|)$. 
This has infinitely many discrete eigenstates described by the modified Bessel function of the second kind
$K_{i\nu}(z)$. The eigenvalues are specified by the {\em pure imaginary zeros\/} of  $K_{i\nu}(g)$ and $K'_{i\nu}(g)$
as a function of the order
with fixed $g>0$.
\end{abstract}

\subsection*{Keywords:}

non-analytic potentials; bound states; modified Bessel function;
pure imaginary orders;
orthogonality theorems; WKB approximation;
associated Hamiltonians;

\subsection*{PACS}

 03.65.Ge Solutions of wave equations: bound states\\
 02.30.Gp Special functions

\section{Introduction}
\label{sec:intro}

This is a second report in the quest of exactly solvable quantum mechanics based on piecewise
analytic potentials.
In this paper we investigate a confining exponential potential $V(x)=g^2\exp(2|x|)$, $g>0$, which 
makes a good contrast to the weak attractive exponential potential $V(x)=-g^2\exp(-|x|)$ explored
in a previous paper \cite{sz1}.
In both cases, the structure of the piecewise analyticity is quite simple.
The exponential potentials have two analytic domains $x<0$ and $x>0$ 
and the eigenfunctions should be matched at the origin $x=0$.
Since the two domains are related by parity $x\leftrightarrow -x$, one needs to consider
only the positive part for the determination of the eigenfunctions.
This simple structure leads to the accessible forms of eigenfunctions; 
(modified) Bessel functions. 
The matching condition at the origin determines the ratio of two independent Bessel function solutions.
The finite norm condition chooses the decreasing member at $x\to\infty$ from the two
independent solutions.
This is carried out by picking up the values of the order of the (modified) Bessel functions
in such a way that they are the zeros of the corresponding Bessel functions at the matching point.
To be more specific,
the zeros $\nu_n$ of $J_\nu(z)$, $J'_\nu(z)$, $K_{i\nu}(z)$, $K'_{i\nu}(z)$ regarded as a function of $\nu$ 
with fixed $z=g$, determine the eigenvalues $E_n=\nu_n^2$ through the matching condition.

This type of solution method \cite{sz1}--\cite{mz1} is markedly different from  that of the other type of 
exactly solvable potentials 
which have various orthogonal polynomials as the
main part of the eigenfunctions.
We do hope these  examples would be followed by many new ones so that a new paradigm 
of exact solvability would be established.

These two examples shed new light on the zeros of (modified) Bessel functions regarded as a
function of the order with fixed argument.
Through present quantum mechanical research, 
rich knowledge of modified Bessel function of the second kind is obtained;
the infinitely many pure imaginary zeros of the order, the orthogonality relations, the asymptotic 
distribution of the zeros, etc.
They are summarised as Theorems and a Conjecture in section 3.

The present paper is organised as follows. In \S\ref{sec:eig} the eigensystems of the 
piecewise analytic exponential potential is derived by imposing matching and finite norm conditions.
A short remark on the infinitely many associated Hamiltonian systems and the corresponding 
orthogonality relations is presented in \S\ref{sec:crum}.
In section three, various results derived in section two are rephrased in mathematical language
by getting rid of quantum mechanical setting.

\section{Quantum Mechanical Setting}
\label{sec:II}

The starting point is  one-dimensional Schr\"odinger equation,
or an eigenvalue problem for a given Hamiltonian $\mathcal{H}$, 
\begin{equation}
\mathcal{H}\psi_n(x)=E_n\psi_n(x),\quad \mathcal{H}\eqdef -\frac{d^2}{dx^2}+V(x).
\label{scheq}
\end{equation}
In this paper we adopt
the non-analytic exponential potential
\begin{equation}
V(x) =g^2\exp(2|x|),\quad x\in(-\infty,\infty), \quad g>0\,.
\label{pot}
\end{equation}
Since the potential grows indefinitely at the boundaries $x=\pm\infty$,
 the system has  infinitely many
bound-states with  energies $E_m=\nu^2_m$, $\nu_m>0$ at $m=0,1,\ldots,$. 
The
corresponding eigenfunctions must be normalizable, $\psi_m(x) \in
L^2(\mathbb{R})$. Since the potential is parity invariant,
$V(-x)=V(x)$, the eigenfunctions are also parity invariant,
\begin{equation}
\psi_m(-x)=(-1)^m\psi_m(x)\,.
\end{equation}
According to the conventional oscillation theorems \cite{Hille} the
subscript $m$ counts the {\em nodes} of the eigenfunctions.
Moreover, we may only consider the (say, positive) half-axis of
$x\ge0$
\begin{equation}
\text{even parity}:\quad\psi'_{2n}(0)=0,\qquad \text{odd
parity}:\quad \psi_{2n+1}(0)=0\,,
 \label{bc}
\end{equation}
{\em i.e.,}  with the eigenfunctions constrained by the parity-dependent
boundary condition at the origin.

\subsection{Eigenfunctions}
\label{sec:eig}

Let us introduce an auxiliary function $\rho(x)$,
\begin{equation}
\rho(x)\eqdef g e^{|x|},\quad
\frac{d\rho(x)}{dx}=\text{sign}(x)\rho(x), \label{rprop}
\end{equation}
which maps $[0,\infty)$ and $(-\infty,0]$ to $[g,\infty)$. With this
function, the Schr\"odinger equation for the discrete spectrum
$E=\nu^2$, $\nu>0$ is rewritten as the equation for Bessel functions,
\begin{equation}
\psi(x)=\phi\bigl(\rho(x)\bigr)\ \Rightarrow
\frac{d^2\phi(\rho)}{d\rho^2}
+\frac1{\rho}\frac{d\phi(\rho)}{d\rho}
-\left(1-\frac{\nu^2}{\rho^2}\right)\phi(\rho)=0.
\end{equation}
The general solutions are obtained as linear combinations of two
types of modified Bessel functions with the {\em pure imaginary order\/} $i\nu$:
\begin{equation*}
I_{i\nu}\bigl(\rho(x)\bigr),\quad I_{-i\nu}\bigl(\rho(x)\bigr).
\end{equation*}
The modified Bessel function of the first kind is defined by \cite{watson}
\begin{equation}
I_{\alpha}(x)=\left(\frac{x}2\right)^{\alpha}\sum_{k=0}^\infty
\frac{\left(\frac{x}2\right)^{2k}}{k!\Gamma(\alpha+k+1)}.
\end{equation}

As real solutions we choose the symmetric and anti-symmetric combinations ($\nu\ge0$),
\begin{align}
I^{(+)}_{i\nu}\bigl(\rho(x)\bigr)&\eqdef \frac12\left[I_{i\nu}\bigl(\rho(x)\bigr)+ I_{-i\nu}\bigl(\rho(x)\bigr)\right],\\
K_{i\nu}\bigl(\rho(x)\bigr)&\eqdef 
\frac{-i\pi}{2\sinh \nu\pi}\left[I_{-i\nu}\bigl(\rho(x)\bigr)- I_{i\nu}\bigl(\rho(x)\bigr)\right],
\quad K_{-i\nu}(\rho)=K_{i\nu}(\rho).
\label{knudef}
\end{align}
Here $K_{\alpha}(z)$ is called modified Bessel function of the second kind, 
which is an entire function of the order $\alpha$ and it satisfies the following relations \cite{watson}: 
\begin{align}
K_{\alpha-1}(z)-K_{\alpha+1}(z)=-\frac{2\alpha}{z}K_{\alpha}(z),\quad
K_{\alpha-1}(z)+K_{\alpha+1}(z)=-2K'_{\alpha}(z).
\label{karel}
\end{align}

Let us look for the eigenfunctions for the even and odd parity as
\begin{align}
\psi^{(e)}(x)=A\, I^{(+)}_{i\nu}\bigl(\rho(x)\bigr)+B\, K_{i\nu}\bigl(\rho(x)\bigr),\quad
A,B\in\mathbb{R},\\
\psi^{(o)}(x)=C\, I^{(+)}_{i\nu}\bigl(\rho(x)\bigr)+D\, K_{i\nu}\bigl(\rho(x)\bigr),\quad
C,D\in\mathbb{R},
\end{align}
These solutions are to be constrained by the appropriate asymptotic
boundary conditions at $x=+\infty$ and by the matching conditions
at the origin \eqref{bc}.

 First let us impose the matching conditions at the origin $x=0$.
Note that, for $\rho(0)=g>0$, the functions $I_{\pm i\nu}(g)$ together with
their derivatives ${I'_{\pm i\nu}}(g)=\left.\frac{d}{d\rho}I_{\pm i\nu}(\rho)\right|_{\rho=g}$ are finite
and non-singular.
Thus, wave functions satisfying the matching conditions \eqref{bc} at the
origin can be easily found. They are
\begin{align}
\psi^{(e)}(x)&=A(\nu,g)\,
I^{(+)}_{i\nu}\bigl(\rho(x)\bigr)+B(\nu,g)\, K_{i\nu}\bigl(\rho(x)\bigr),\\
&A(\nu,g)\eqdef -K_{i\nu}'(g),\quad 
B(\nu,g)\eqdef
{I^{(+)}_{i\nu}}'(g),
\label{ecomb}\\
\psi^{(o)}(x)&=C(\nu,g)\,
 I^{(+)}_{i\nu}\bigl(\rho(x)\bigr)+D(\nu,g)\, K_{i\nu}\bigl(\rho(x)\bigr),\\
&C(\nu,g)\eqdef -K_{i\nu}(g),\quad
D(\nu,g)\eqdef I^{(+)}_{i\nu}(g). 
\label{ocomb}
\end{align}

The finite norm condition or the asymptotic condition at $x\to+\infty$, ($\rho\to+\infty$) is easily imposed.
The  modified Bessel functions have the following asymptotic behaviour \cite{watson}:
\begin{align}
I_{\alpha}(x)&\sim \frac{e^x}{\sqrt{2\pi x}}\sum_{n=0}^\infty \frac{(-1)^n(\alpha,n)}{(2x)^n}+
\frac{e^{-x+(\alpha+\tfrac12)\pi i}}{\sqrt{2\pi x}}\sum_{n=0}^\infty \frac{(\alpha,n)}{(2x)^n},\quad
(-\pi/2<\text{arg}\, x<3\pi/2),\\[4pt]
I_{\alpha}(x)&\sim \frac{e^x}{\sqrt{2\pi x}}\sum_{n=0}^\infty \frac{(-1)^n(\alpha,n)}{(2x)^n}+
\frac{e^{-x-(\alpha+\tfrac12)\pi i}}{\sqrt{2\pi x}}\sum_{n=0}^\infty \frac{(\alpha,n)}{(2x)^n},\quad
 (-3\pi/2<\text{arg}\, x<\pi/2),\\[4pt]
K_{\alpha}(x)&\sim \sqrt{\frac{\pi}{2x}}\,e^{-x}\left[1+\sum_{n=1}^\infty\frac{(\alpha,n)}{(2x)^n}\right],
\qquad \qquad \qquad (|\text{arg}\, x|<3\pi/2),
\label{Kalaym}
\end{align}
in which $(\alpha,n)$ is defined by 
\begin{equation*}
(\alpha,n)\eqdef (-1)^n(\alpha+1/2)_n(-\alpha+1/2)_n/n!.
\end{equation*}
Here $(a)_n\eqdef\prod_{k=1}^n(a+k-1)$ is the shifted factorial or the so-called Pochhammer's symbol.

The eigenvalues are selected by requiring the  finite norm condition.
That is, the coefficients $A(\nu,g)$ \eqref{ecomb} and $C(\nu,g)$ \eqref{ocomb} of the divergent term 
$I^{(+)}_{i\nu}\bigl(\rho(x)\bigr)$ should vanish:
\begin{align}
\text{even:}&\quad
K'_{i\nu_{2n}}(g)=0,\quad \quad
n=0,1,\ldots,
\label{even}\\
\text{odd:}&\quad K_{i\nu_{2n+1}}(g)=0,\quad \, n=0,1,\ldots\,.
\label{odd}
\end{align}
Thus, one obtains the {\em eigenfunctions} 
\begin{align}
\psi_{2n}(x)&=K_{i\nu_{2n}}\bigl(\rho(x)\bigr)=\left\{
\begin{array}{cc}
K_{i\nu_{2n}}\bigl(g\,e^{x})  &   x\ge0   \\[2pt]
K_{i\nu_{2n}}\bigl(g\,e^{-x})  &   x\le0
\end{array}
\right., \  \ E_{2n}=\nu_{2n}^2, \quad n=0,1,\ldots,
\label{evenfun}\\[4pt]
\psi_{2n+1}(x)&=\text{sign}(x)K_{i\nu_{2n+1}}\bigl(\rho(x)\bigr)=\left\{
\begin{array}{cc}
K_{i\nu_{2n+1}}\bigl(g\,e^{x})  &   x\ge0   \\[2pt]
-K_{i\nu_{2n+1}}\bigl(g\,e^{-x})  &   x\le0
\end{array}
\right. ,\\
&\hspace{75mm} E_{2n+1}=\nu_{2n+1}^2, \quad n=0,1,\ldots,
\label{oddfun}\\
&  g^2<E_0<E_1<E_2<\cdots \quad \Longleftrightarrow \quad
g<\nu_0<\nu_1<\nu_2<\cdots . 
\label{e-kappa}
\end{align}

In this manner we have established that our ``confining non-analytic
exponential  potential'' (\ref{pot}) is {\em exactly solvable}.
Moreover, the compact form of the solutions enables us  to
verify easily the mutual orthogonality between the even and odd
eigenfunctions. The evaluation of the normalisation
constants $h_{2n}$, $h_{2n+1}$ in the relations
\begin{align}
\int_0^{\infty}K_{i\nu_{2n}}(g\,e^{x})
K_{i\nu_{2m}}(g\,e^{x})\,dx&=
h_{2n}\delta_{n\,m},
\label{orteven}\\
\int_0^{\infty}K_{i\nu_{2n+1}}(g\,e^{x})
K_{i\nu_{2m+1}}(g\,e^{x})dx&=
h_{2n+1}\delta_{n\,m}, \label{ortodd}
\end{align}
is left to the readers.

It is important to note that the functions $K_{i\nu}(g)$, $K'_{i\nu}(g)$ as functions of $\nu$ for
fixed $g>0$ are rapidly decreasing partly due to the factor $1/\sinh\nu\pi$ in \eqref{knudef}, 
see \eqref{erd}. This makes numerical determination of the eigenvalues rather delicate. 
An approximate formula of the eigenvalues $E_n=\nu_n^2$  in the asymptotic region $n\gg1$
can be obtained in terms of the WKB approximation or the Bohr-Sommerfeld quantum condition
formula \eqref{wkb} described in Conjecture\,\ref{wkbconj}.

\subsection{Associated Hamiltonians}
\label{sec:crum}

According to Crum \cite{crum}, to a one-dimensional Hamiltonian
$\mathcal{H}=\mathcal{H}^{[0]}$ with the eigensystem
$\{E_n,\psi_n(x)\}$, $n=0,1,\ldots$, a sequence of {\em
iso-spectral} Hamiltonian systems $\mathcal{H}^{[L]}$ $L=1,2,\ldots$,
is associated:
\begin{align}
\mathcal{H}^{[L]}\psi_n^{[L]}(x)&=E_n\psi_n^{[L]}(x),
\quad n=L,L+1,\ldots,\\
\mathcal{H}^{[L]}&\eqdef\mathcal{H}^{[0]}
-2\partial_x^2\log\left|\text{W}[\psi_0,\psi_1,\ldots,\psi_{L-1}](x)\right|,\\
\psi_n^{[L]}(x)&\eqdef\frac{\text{W}[\psi_0,\psi_1,\ldots,
\psi_{L-1},\psi_n](x)}{\text{W}[\psi_0,\psi_1,\ldots,\psi_{L-1}](x)},
\label{wronM}
\end{align}
in which the Wronskian of $n$-functions $\{f_1,\ldots,f_n\}$ is
defined by formula
\begin{align}
&\text{W}\,[f_1,\ldots,f_n](x)
  \eqdef\det\Bigl(\frac{d^{j-1}f_k(x)}{dx^{j-1}}\Bigr)_{1\leq j,k\leq n}.
  \label{wron}
  \end{align}
This result is obtained from a multiple application of the Darboux
transformations \cite{darboux}. 

Let us apply Crum's sequence to the present Hamiltonian \eqref{scheq}, \eqref{pot}, 
\eqref{evenfun}, \eqref{oddfun}.
The infinitely many associated Hamiltonian systems
$\mathcal{H}^{[L]}$ with $L=1,2,\ldots$  are
all {\em exactly solvable}. It is easy to see that the systems are
parity invariant:
\begin{align}
V^{[L]}(x)&\eqdef V(x)-2\partial_x^2\log
\left|\text{W}[\psi_0,\psi_1,\ldots,\psi_{L-1}](x)\right|,
\quad V^{[L]}(-x)=V^{[L]}(x),\\
\psi_n^{[L]}(-x)&=(-1)^{L+n}\psi_n^{[L]}(x).
\end{align}
Because of the parity, the orthogonality relations among the even
and odd eigenfunctions are trivial and those even-even and odd-odd
\begin{align}
\delta_{n\,m}\propto
(\psi_n^{[L]},\psi_m^{[L]})=\int_{-\infty}^{\infty}
\psi_n^{[L]}(x)\psi_m^{[L]}(x)dx
\end{align}
can be rewritten as those on the positive $x$-axis
\begin{align}
\delta_{n\,m}&\propto \int_{0}^{\infty}
\psi_{2n}^{[L]}(x)\psi_{2m}^{[L]}(x)dx,
\label{ortMeven}\\
\delta_{n\,m}&\propto \int_{0}^{\infty}
\psi_{2n+1}^{[L]}(x)\psi_{2m+1}^{[L]}(x)dx. \label{ortModd}
\end{align}

By using known properties of the Wronskians \cite{sz1},
we can reduce the Wronskians
of $\{\psi_n(x)\}$ in  \eqref{wronM}  to the Wronskians of the
modified Bessel functions of the second kind, $\{K_{i\nu_n}(\rho)\}$. 
This makes the actual evaluation much simpler, for
example, we obtain:
\begin{align}
\text{W}[\psi_0,\psi_n](x)&=\left(\text{sign}(x)\right)^{1+n}\rho
\cdot \text{W}[K_{i\nu_0}(\rho),K_{i\nu_n}(\rho)](\rho),\n
\text{W}[\psi_0,\psi_1,\psi_n](x)&=\left(\text{sign}(x)\right)^{2+n}
\rho^3\!\cdot
\text{W}[K_{i\nu_0}(\rho),K_{i\nu_1}(\rho),K_{i\nu_n}(\rho)](\rho),\n
\text{W}[\psi_0,\psi_1,\ldots,\psi_{L-1},\psi_n](x)&=
\left(\text{sign}(x)\right)^{L+n}\rho^{L(L+1)/2}\cdot\n
&\quad\times
\text{W}[K_{i\nu_0}(\rho),K_{i\nu_1}(\rho),\ldots,K_{i\nu_{L-1}}(\rho),
K_{i\nu_n}(\rho)](\rho).
\end{align}

It is straightforward to evaluate $V^{[1]}(x)$ explicitly. It is not of the form 
\begin{align*}
f(g)^2e^{2|x|},
\end{align*}
with a certain function $f(g)$ of the parameter $g$, up to an additive constant. 
That is, the system is
{\em not shape invariant}.

%
%
%
\section{Mathematical Reinterpretation}

The results obtained in the previous section can be stated as 
various Theorems on modified Bessel functions of the second kind.

\begin{theo}{\bf Pure imaginary zeros}\label{theo:one}\quad
When modified Bessel functions of the second kind $K_{\alpha}(x)$, $\frac{d}{dx}K_{\alpha}(x)$
are regarded as functions of the order $\alpha$ for fixed $x>0$,
they have infinitely many pure imaginary zeros:
\begin{align} 
   \frac{dK_{\pm i\lambda_j}(x)}{dx}&=0, \qquad 0<x<\lambda_0<\lambda_1<\lambda_2<\ \cdots, 
   \label{lams}\\
   K_{\pm i\mu_j}(x)&=0, \qquad 0<x<\mu_0<\mu_1<\mu_2<\ \cdots.
   \label{mus}
\end{align}
They are interlaced by the oscillation theorem:
\begin{equation}
0<x<\lambda_0<\mu_0<\lambda_1<\mu_1<\lambda_2<\mu_2<\,\cdots .
\end{equation}
\end{theo}
Since the discrete eigenvalues of one dimensional quantum mechanics are always simple, 
all these zeros are also simple.
\begin{theo}{\bf Orthogonality relation \I}\label{theo:two}\quad 
The Bessel function of the second kind with the above pure imaginary orders $\{i\lambda_j\}$ \eqref{lams},
$\{i\mu_j\}$ \eqref{mus} satisfy the following orthogonality relations \rm{(}$x>0$\rm{)}:
\begin{align}
\text{even}:& \quad
\int_x^{\infty}K_{i\lambda_j}(\rho)K_{i\lambda_k}(\rho)\frac{d\rho}{\rho}=0,\qquad
j\neq k,
\label{theo1even}\\
\text{odd}:& \quad
\int_x^{\infty}K_{i\mu_j}(\rho)K_{i\mu_k}(\rho)\frac{d\rho}{\rho}=0,\qquad
j\neq k. \label{theo1odd}
\end{align}
\end{theo}
Let us denote these two types of zeros by one consecutive sequence
($\{\nu_j\}$):
\begin{align*}
\nu_0\equiv \lambda_0,\ \nu_1\equiv \mu_0,\ \nu_2\equiv \lambda_1,\
\nu_3\equiv \mu_1,\ldots,.
\end{align*}

The  orthogonality relations of the eigenfunctions
\eqref{ortMeven}--\eqref{ortModd} of the $L$-th associated
Hamiltonian system can be stated as\
\begin{theo}{\bf Orthogonality relation \II}\label{theo:thr}
\begin{align}
&{\rm even}:\int_x^{\infty}\frac{\text{W}[K_{i\nu_0},\ldots,K_{i\nu_{L-1}},
K_{i\nu_{2n}}](\rho) \text{W}[K_{i\nu_0},\ldots,K_{i\nu_{L-1}},
K_{i\nu_{2m}}](\rho)}
{\left(\text{W}[K_{i\nu_0},\ldots,K_{i\nu_{L-1}}](\rho)\right)^2}\rho^{2L-1}
d\rho=0,\n
& \hspace{135mm} n\neq m,
\label{theo2even}\\
&{\rm odd}:\int_x^{\infty}\frac{\text{W}[K_{i\nu_0},\ldots,K_{i\nu_{L-1}},
K_{i\nu_{2n+1}}](\rho) \text{W}[K_{i\nu_0},\ldots,K_{i\nu_{L-1}},
K_{i\nu_{2m+1}}](\rho)}
{\left(\text{W}[K_{i\nu_0},\ldots,K_{i\nu_{L-1}}](\rho)\right)^2}\rho^{2L-1}
d\rho=0,\n 
&\hspace{135mm}
n\neq m. \label{theo2odd}
\end{align}
\end{theo}
Theorem\,\ref{theo:thr} is the special case ($L=0$) of Theorem\,\ref{theo:two}.

Orthogonality relation I can be derived directly based on an indefinite integration formula 
of Bessel functions in Watson's textbook \cite{watson} (formula (13) on page 135, \S5.11)
\begin{align} 
 \int^z \mathcal{C}_{\mu}(kz)\overline{\mathcal{C}}_{\nu}(kz)\frac{dz}{z}&=
-\frac{kz\left\{\mathcal{C}_{\mu+1}(kz)\overline{\mathcal{C}}_{\nu}(kz)
-\mathcal{C}_{\mu}(kz)\overline{\mathcal{C}}_{\nu+1}(kz)\right\}}{\mu^2-\nu^2} \nonumber\\[-4pt]
&\quad \mbox{}+\frac{\mathcal{C}_{\mu}(kz)\overline{\mathcal{C}}_{\nu}(kz)}{\mu+\nu},
\end{align}
in which $\mathcal{C}_{\mu}$ and  $\overline{\mathcal{C}}_{\nu}$ are two arbitrary cylinder functions.
For the odd part \eqref{theo1odd}  
one puts  $k=1$, $\mathcal{C}_{\mu}(z)\to K_{i\mu}(z)$, $\overline{\mathcal{C}}_{\nu}(z)\to K_{i\nu}(z)$
and integrate over $[z,\infty)$.
The even part \eqref{theo1even}  can also be obtained in the same manner by noting,
in terms of \eqref{karel}, $K'_{i\mu}(z)=0$\ 
$\Rightarrow -K_{i\mu-1}(z)=K_{i\mu+1}(z)=\frac{i\mu}{z}K_{i\mu}(z)$.

As for the asymptotic distribution of the pure imaginary zeros $\{\nu_n\}$, $n\gg1$, 
we can make a conjecture based on the WKB approximation or the so-called Bohr-Sommerfeld 
quantum condition
$\oint p(x)dx=2\pi (n+\tfrac12)$. Here $p(x)$ is the momentum at $x$. For $x\ge0$, it is determined by the
energy conservation $p^2+g^2\,e^{2x}=E_n=\nu_n^2$. Based on the formula
\begin{equation}
4\int_0^{\log(\nu_n/g)}\!\!\sqrt{\nu_n^2-g^2\,e^{2x}}\,dx=2\pi (n+\tfrac12),
\end{equation}
we arrive at the following
\begin{conj}{\bf Asymptotic distribution of the pure imaginary zeros}\label{wkbconj}\\
The asymptotic location of the $n$-th pure imaginary zero $i\nu_n$ of combined  $K_{\alpha}(x)$ and $K'_{\alpha}(x)$, $x>0$,
is given by
\begin{equation}
\nu_n\,{\rm arccosh}\biggl(\frac{\nu_n}{x}\biggr)-\sqrt{\nu_n^2-x^2}=\frac{(n+\tfrac12)\pi}{2},\qquad n\gg1.
\label{wkb}
\end{equation}
The $n$-th pure imaginary zero of $K_{\alpha}(x)$ only or of $K'_{\alpha}(x)$ only obeys the same rule but $n$ in the r.h.s. should be replaced by $2n$.
\end{conj}
This is consistent with the main term of the asymptotic expression of $K_{i\nu}(x)$, $\nu>x>0$ (\cite{erdelyi} \S7.13.2 formula (19)),
\begin{align} 
K_{i\nu}(x)&=\sqrt{2}(\nu^2-x^2)^{-\tfrac14}e^{-\nu\pi/2}\left[\sqrt\pi\sin\left(\nu\,{\rm arccosh}\biggl(\frac{\nu}{x}\biggr)-\sqrt{\nu^2-x^2}+\frac{\pi}4\right)\right.\n
&\hspace{60mm}\left.
+O\biggl(\frac1{\sqrt{\nu^2-x^2}}\biggr)\right].
\label{erd}
\end{align}

\bigskip
After completing this work, we are informed of a work \cite{mz2} discussing the same problem from a
rather different angle.

\subsection*{Acknowledgements}

The author thanks Milosh Znojil for many interesting discussions on exact solvability.
He also thanks Jeffrey Lagarias for enlightening communication.




\end{document}